\documentclass{article}
\usepackage[preprint]{spconfa4}
\usepackage{amsmath,graphicx}

\usepackage{booktabs}
\usepackage{url}
\usepackage{balance}
\usepackage{makecell}
\usepackage{cite}

\copyrightnotice{%
\parbox[b]{\textwidth}{\scriptsize
\copyright\ 2026 IEEE. Personal use of this material is permitted. Permission from IEEE must be obtained for all other uses, in any current or future media, including reprinting/republishing this material for advertising or promotional purposes, creating new collective works, for resale or redistribution to servers or lists, or reuse of any copyrighted component of this work in other works. Accepted for publication in the Proceedings of the 19th International Workshop on Acoustic Signal Enhancement (IWAENC 2026).}}

\title{PathRIR: Physics-Guided Acoustic Path Selection and Late-Tail Compensation for Fast Room Impulse Response Simulation}

\name{\begin{tabular}{@{}c@{}}
Shaoheng Xu$^{1,*,\dagger}$, Chunyi Sun$^{1,*}$, Jihui (Aimee) Zhang$^{2,1}$, Amy Bastine$^{1}$,\\
Prasanga N. Samarasinghe$^{1}$, Thushara D. Abhayapala$^{1}$
\end{tabular}
\thanks{$^{*}$These authors contributed equally. $^{\dagger}$The corresponding author.\\
Code and data: \protect\url{https://github.com/ShaoHenry/PathRIR}}}
\address{$^1$The Australian National University, Australia\\
$^2$The University of Queensland, Australia}

\begin{document}
\ninept
\maketitle

\begin{abstract}
Image-source-method (ISM)-based room impulse response (RIR) simulation is a useful and physically interpretable tool for acoustic scene modeling, but full-order ISM becomes computationally expensive as the reflection order and room complexity increase. We propose a physics-guided framework for fast RIR simulation that preserves the geometric structure of ISM while learning to retain only acoustically important image-source paths during online traversal. To recover energy removed by pruning, the proposed \textit{PathRIR} uses a lightweight compensation multilayer perceptron to predict the missing late-tail energy envelope and generate a compensation tail whose energy follows that envelope. Experiments on irregular 3D rooms show that \textit{PathRIR} reduces image-source computation and improves runtime efficiency over a full-order ISM simulator, while achieving low waveform- and decay-related errors. Ablation results show that adding the compensation tail improves waveform fidelity and reduces energy-decay-curve error, reverberation-time error, and direct-to-reverberant-ratio error, with modest runtime overhead.
\end{abstract}


% ============================================================= %
% ============================================================= %
% ============================================================= %

\vspace{-0.4cm}
\section{Introduction}
\label{sec:intro}

Room impulse responses (RIRs) are fundamental descriptors of indoor acoustic propagation and provide controlled acoustic data for speech and audio systems. An RIR captures the transfer path between a sound source and a receiver, including the direct sound, early reflections, and late reverberation, and is widely used in speech enhancement, dereverberation, source localization, spatial audio, and data augmentation~\cite{ASR-2017,RIR-SSL,dereverb-2010,RIR-ISO,soundspaces,AuralizationBook}. As speech and audio models continue to advance, the demand for large, diverse, and controllable acoustic datasets is increasing, while dense real-world RIR measurements remain costly, time-consuming, and environment-specific~\cite{rir-former,diffusionrir}.

Existing approaches to RIR simulation include wave-based solvers, geometric acoustics methods, and hybrid systems. Wave-based solvers can accurately model acoustic wave phenomena, but are computationally demanding~\cite{ard,diffuse-acoustic}. Geometric methods, such as the image source method (ISM), ray tracing, and diffuse acoustic simulation~\cite{ga-overview,ism,rir-g,pyroomacoustics,ism-polyhedra,diffuse-acoustic}, are widely used because they offer efficient and physically interpretable approximations of acoustic propagation. In particular, ISM represents sound propagation as image-source paths, making delay, attenuation, and direction of arrival explicit or directly derivable. It is used in tools such as Habets' RIR Generator~\cite{rir-g}, Pyroomacoustics~\cite{pyroomacoustics}, and gpuRIR~\cite{gpuRIR}, and has been extended to directional-transducer simulation using spherical harmonics~\cite{gism}. However, this path-level interpretability comes with a cost: in shoebox rooms, the number of image sources grows cubically with reflection order, and high-order simulation requires many reflected contributions to reproduce the late reverberant tail. Although GPU implementations~\cite{gpuRIR}, ray/path-tracing simulators~\cite{pyroomacoustics,diffuse-acoustic}, and hybrid systems such as Treble~\cite{treble} improve efficiency or late-field modeling, large-scale high-order RIR generation can still be expensive for online training and rendering.

Several recent methods improve efficiency or realism by approximating, transforming, or learning the RIR generation process. FRA-RIR~\cite{fra-rir} and FRAM-RIR~\cite{fram-rir} accelerate ISM-style simulation by approximating virtual-source propagation structure, while neural generators and translation-based methods such as FAST-RIR~\cite{fast-rir}, MESH2IR~\cite{mesh2ir}, IR-GAN~\cite{ir-gan}, TS-RIR~\cite{ts-rir}, and audio-visual or neural-field approaches~\cite{av-rir,nacf,neraf} learn to synthesize or transform RIRs from room parameters, meshes, visual cues, spatial coordinates, acoustic parameters, or existing synthetic RIRs. These methods enable fast or more realistic RIR generation, but they usually generate the final RIR directly, which can reduce waveform fidelity and weaken the path-level structure of geometric simulation: individual reflections are not explicitly selected, evaluated, or exposed as controllable physical components.

This paper explores a different trade-off from direct neural RIR generation. Rather than replacing the geometric simulator, we propose \textit{PathRIR}, a physics-guided neural acceleration framework that learns to prune ISM traversal and compensate for the resulting late-tail energy loss. Many candidate paths contribute little individually, especially at high reflection orders, but discarded high-order paths can still collectively shape the late reverberant decay. Therefore, effective acceleration should prune negligible branches while preserving the aggregate late-tail energy that would otherwise be lost. In \textit{PathRIR}, a lightweight Pruning-MLP predicts whether each image-source subtree should continue to be expanded, while retained paths are still computed by ISM so their delays, attenuation, and reflection orders remain physically interpretable. A lightweight Compensation-MLP predicts the missing late-tail energy envelope used to shape a stochastic compensation tail~\cite{diffuse-rir}. Our contributions are threefold: (1) a neural image-source path-selection framework that accelerates geometric RIR simulation while retaining explicit, physically interpretable path contributions; (2) subtree-level supervision that estimates the aggregate importance of an image-source branch rather than only an individual reflection; and (3) residual late-tail compensation that restores energy decay after aggressive pruning. We evaluate the resulting speed--accuracy trade-off against full-order ISM simulation and neural RIR generation across irregular 3D room geometries.

% ============================================================= %
% ============================================================= %
% ============================================================= %

\section{Problem Formulation}
\label{sec:formulation}
\vspace{-0.1cm}

We consider RIR simulation in a known room geometry, including irregular room shapes, with one fixed source and $M$ microphones, indexed by $m=1,\ldots,M$. Given this source--room configuration, ISM constructs a set of image-source nodes $\mathcal{V}$, where each node $v\in\mathcal{V}$ represents one possible direct or reflected sound path. The set $\mathcal{V}$ includes the order-zero direct path and all image-source nodes up to a chosen maximum reflection order $O_{\max}$. The full-order RIR at microphone $m$ is written as
\begin{equation}
\vspace{-0.05cm}
    h_m(t)
    =
    \sum_{v \in \mathcal{V}} h_{v,m}(t),
\label{eq:full_rir}
\vspace{-0.05cm}
\end{equation}
where $h_{v,m}(t)$ denotes the delayed and attenuated contribution of node $v$ to microphone $m$. Our goal is to accelerate ISM by computing only the image-source paths that are acoustically important. During ISM traversal, removing one node also removes all paths that would be generated from it. Therefore, the pruning decision should consider the total contribution of the whole subtree, rather than only the energy of a single node. Let $\mathcal{T}(v)$ denote the subtree rooted at node $v$. We define the normalized subtree importance as
\begin{equation}
    I(v)
    =
    \frac{
    \sum_{m=1}^{M}
    \left\|
    \sum_{u \in \mathcal{T}(v)} h_{u,m}(t)
    \right\|_2^2
    }{
    \sum_{m=1}^{M}
    \left\|
    h_m(t)
    \right\|_2^2
    }.
\label{eq:subtree_importance}
\end{equation}
This quantity measures how much acoustic energy would be lost if node $v$ and its descendants were pruned.

A learned pruning policy selects a retained node set $\mathcal{A}_\theta \subseteq \mathcal{V}$ during online image-source expansion. Using only these retained nodes, the pruned ISM response is
\begin{equation}
    \widetilde{h}_m(t)
    =
    \sum_{v \in \mathcal{A}_\theta}
    h_{v,m}(t).
\label{eq:pruned_rir}
\end{equation}
This keeps the analytic image-source contributions for the selected paths, while reducing the number of nodes that need to be expanded and accumulated.

However, aggressive pruning can remove many weak high-order paths that collectively shape the late reverberant tail. To compensate for this missing energy, we add a statistical residual term $\eta_m(t)$, generated from a predicted residual energy envelope:
\begin{equation}
    \widehat{h}_m(t)
    =
    \widetilde{h}_m(t) + \eta_m(t).
\label{eq:compensated_rir}
\end{equation}
The objective is to reduce simulation cost while keeping the compensated response $\widehat{h}_m(t)$ close to the full-order response $h_m(t)$ in terms of both waveform shape and energy-decay properties.

% ============================================================= %
% ============================================================= %
% ============================================================= %

\section{Proposed Method}
\label{sec:method}

\begin{figure*}[t!]
  \centering
  \includegraphics[width=1\linewidth]{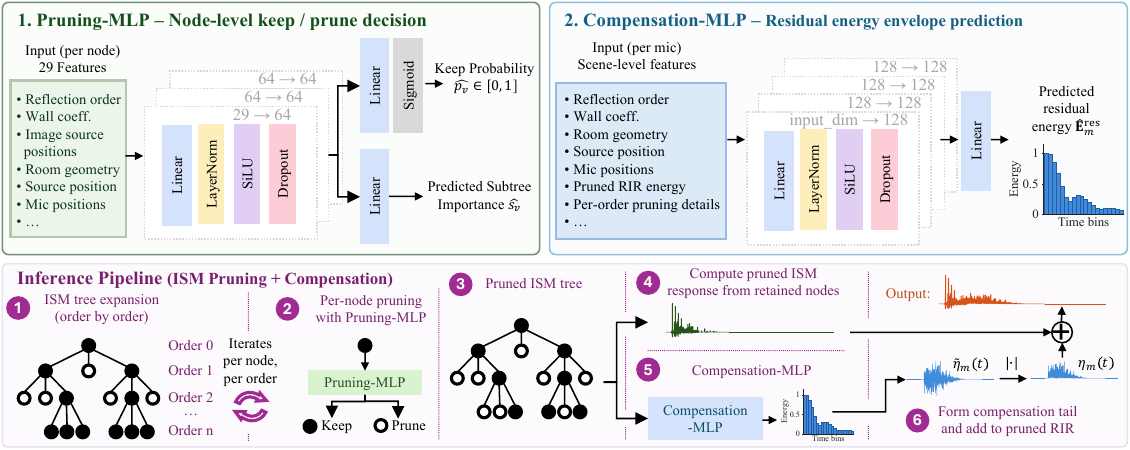}
  \vspace{-0.35cm}
  \caption{\textbf{Overview of the proposed \textit{PathRIR} pipeline.} During order-wise ISM expansion, the Pruning-MLP makes a keep/prune decision for each candidate image-source node. After reaching $O_{\max}$, the pruned RIR is synthesized from the retained nodes, while the Compensation-MLP predicts the missing residual-energy envelope used to scale zero-mean Gaussian noise. The absolute value of the scaled noise forms the compensation tail, which is added to the pruned RIR to produce the final \textit{PathRIR} output.}
  \label{fig:pipeline}
  \vspace{-0.2cm}
\end{figure*}

\textit{PathRIR} combines an analytic ISM simulator with two lightweight learned modules, as shown in Fig.~\ref{fig:pipeline}: an image-source branch selector, or Pruning-MLP, and a statistical late-tail compensator, or Compensation-MLP. The Pruning-MLP keeps only important image-source branches during simulation, while the Compensation-MLP estimates and restores the missing late reverberant energy caused by pruning.

% ------------------------------------------------------------ %

\subsection{Pruning-MLP label generation}

For each training room, we run a full-order ISM simulation to obtain the image-source tree. For each node $v$ in the tree, we compute its subtree importance $I(v)$ using Eq.~\eqref{eq:subtree_importance} and assign the binary keep target
$y_v=\mathbf{1}[I(v)\geq\epsilon_{\mathrm{imp}}]\in\{0,1\}$,
where $\epsilon_{\mathrm{imp}}$ is a predefined importance threshold. A node is labeled important if the aggregate contribution of its subtree is significant. This subtree-level supervision encourages the Pruning-MLP to keep branches that are acoustically important as a whole, rather than judging each node only by its individual energy.

% ------------------------------------------------------------ %

\subsection{Pruning-MLP training}

For each node $v$, we construct a feature vector $\mathbf{f}_v$ using physical information about the node and the room, such as reflection order, image-source location, parent node, accumulated attenuation, source--microphone distance and delay, and simple room-geometry features. The Pruning-MLP predicts a keep probability and a subtree-importance score: $ (\widehat{p}_v, \widehat{s}_v)=f_\theta(\mathbf{f}_v)$, where $\widehat{p}_v\in[0,1]$ is the predicted probability that node $v$ should be kept, and $\widehat{s}_v$ is the predicted subtree-importance score. The target score is $s_v=\log_{10}(I(v)+\epsilon)$, where $\epsilon>0$ is a small constant used to avoid $\log(0)$. Overall, the Pruning-MLP is trained with
\begin{equation}
    \mathcal{L}_{\mathrm{prune}}
    =
    \mathcal{L}_{\mathrm{WBCE}}(\widehat{p}_v,y_v)
    +
    \lambda_{\mathrm{reg}}
    \mathcal{L}_{\mathrm{reg}}(\widehat{s}_v,s_v),
\end{equation}
where $\mathcal{L}_{\mathrm{WBCE}}$ is a weighted binary cross-entropy loss with a larger penalty for wrongly pruning important nodes. The second term is a smooth-$\ell_1$ regression loss for predicting the importance score $s_v$.

% ------------------------------------------------------------ %

\subsection{Pruning-MLP inference}

During inference, the ISM tree is expanded iteratively, order by order. At each reflection order, the Pruning-MLP evaluates each candidate image-source node using $\widehat{p}_v$ and $\widehat{s}_v$. If a node is pruned, its descendants are not generated, reducing both image-source tree expansion and RIR accumulation cost.

Let $o_v$ be the reflection order of node $v$, and let $\mathcal{C}_o$ be the candidate nodes with $o_v=o$. We first select nodes whose predicted keep probability exceeds a threshold $\tau$: $\mathcal{K}^{\mathrm{raw}}_o = \{v \in \mathcal{C}_o : \widehat{p}_v \geq \tau\}.$

A fixed threshold may prune too many high-order nodes and make the late reverberant tail decay too quickly. We therefore use an order-wise budget: all early nodes with $o\leq O_{\mathrm{early}}$ are kept, while for later orders $o > O_{\mathrm{early}}$,
\begin{equation}
\begin{aligned}
    b^{\min}_o
    &=
    \max\!\left(
    \left\lceil r_{\min}\times|\mathcal{C}_o|\right\rceil,
    n_{\min}
    \right), \\
    b^{\max}_o
    &=
    \max\!\left(
    b^{\min}_o,
    \left\lceil r_{\max}\times|\mathcal{C}_o|\right\rceil
    \right), \\
    q_o
    &=
    \min\!\left(
    \max\!\left(
    |\mathcal{K}^{\mathrm{raw}}_o|,
    b^{\min}_o
    \right),
    b^{\max}_o
    \right).
\end{aligned}
\end{equation}
Here, $|\mathcal{C}_o|$ is the number of candidate nodes at reflection order $o$, $r_{\min}$ and $r_{\max}$ are predefined minimum and maximum keep rates, $n_{\min}$ is the predefined minimum number of nodes to keep, and $\lceil\cdot\rceil$ denotes rounding up to the nearest integer. The final retained set $\mathcal{K}_o$ contains the top-$q_o$ candidate nodes ranked by the predicted importance score $\widehat{s}_v$. Repeating this process up to order $O_{\max}$ gives the final retained node set $\mathcal{A}_\theta=\bigcup_{o=0}^{O_{\max}}\mathcal{K}_o$, from which the pruned RIR $\widetilde{h}_m(t)$ is computed as in Eq.~\eqref{eq:pruned_rir}.

% ------------------------------------------------------------ %

\subsection{Compensation-MLP training and inference}

Pruning can remove many weak high-order reflections that are small individually but important in total for late reverberant decay. The Compensation-MLP restores this missing energy by predicting the energy envelope of the residual left out by pruning. For microphone $m$, we define the residual signal as $e_m(t)=h_m(t)-\widetilde{h}_m(t)$. To represent its energy envelope, we divide the residual into $B$ time bins. Let $\mathcal{B}_b$ denote samples in bin $b$; the target residual energy is
\begin{equation}
    E^{\mathrm{res}}_{m,b}
    =
    \sum_{t \in \mathcal{B}_b} e_m^2(t),
    \quad b=1,\ldots,B.
\end{equation}
The vector $\mathbf{E}^{\mathrm{res}}_m=[E^{\mathrm{res}}_{m,1},\ldots,E^{\mathrm{res}}_{m,B}]$ is the target envelope predicted by the Compensation-MLP. Given an input vector $\mathbf{z}_m$ containing room geometry, source--microphone configuration, pruning information, and pruned-RIR energy features, the Compensation-MLP predicts $\widehat{\mathbf{E}}^{\mathrm{res}}_m = g_\psi(\mathbf{z}_m)$. It is trained with
\begin{equation}
    \mathcal{L}_{\mathrm{comp}}
    =
    \mathcal{L}_{\mathrm{bin}}(
    \widehat{\mathbf{E}}^{\mathrm{res}}_m,
    \mathbf{E}^{\mathrm{res}}_m)
    +
    \lambda_{\mathrm{edc}}
    \mathcal{L}_{\mathrm{EDC}}(
    \widehat{\mathbf{E}}^{\mathrm{res}}_m,
    \mathbf{E}^{\mathrm{res}}_m),
\end{equation}
where $\mathcal{L}_{\mathrm{bin}}$ measures bin-wise residual-energy error, and $\mathcal{L}_{\mathrm{EDC}}$ measures the corresponding energy-decay-curve error.

During inference, after a compensation start time $t_{\mathrm{comp}}$, we generate zero-mean Gaussian noise~\cite{diffuse-rir} and scale it within each time bin to match $\widehat{\mathbf{E}}^{\mathrm{res}}_m$, obtaining $\widetilde{\eta}_m(t)$. We then take its absolute value to form the compensation tail, $\eta_m(t)=|\widetilde{\eta}_m(t)|$, without changing the energy assigned to each time bin. The compensation tail is added to the pruned RIR as in Eq.~\eqref{eq:compensated_rir}, restoring the missing late-tail energy decay rather than the exact residual waveform, while the retained paths remain analytic ISM contributions.

% ============================================================= %
% ============================================================= %
% ============================================================= %

\section{Experiments}
\label{sec:experiments}

\subsection{Experiment setup}
\label{ssec:experiment setup}

We evaluate \textit{PathRIR} (Fig.~\ref{fig:pipeline}) on Monte Carlo-generated 3D irregular rooms. Each room is obtained by extruding a random 2D polygon with 5--10 vertices~\cite{pyroomacoustics, ism-polyhedra}. The floor-plan width and length are sampled in $[3,12]$~m, and the height is sampled in $[2.2,4.5]$~m. Wall, floor, and ceiling absorption coefficients are randomly sampled within $[0.03,0.70]$, producing rooms with $T_{60}$ values in the range $[0.109,0.645]$~s. For each room, one source and two microphones are randomly placed inside the room with a minimum source--microphone distance of 0.75~m. We generate $1000$ rooms for training and $20$ held-out rooms for testing. All RIRs are simulated up to $O_{\max}=10$, sampled at $f_s=8$~kHz, and truncated to 0.5~s. The Pruning-MLP is trained with $\lambda_{\mathrm{reg}}=0.25$ and $\epsilon_{\mathrm{imp}}=10^{-4}$. The Compensation-MLP is trained with $\lambda_{\mathrm{edc}}=0.5$. During inference, \textit{PathRIR} uses threshold $\tau=0.5$, keeps all nodes up to order $O_{\mathrm{early}}=1$, and limits the number of retained nodes at each later order using $r_{\min}=0.2$, $r_{\max}=0.5$, and $n_{\min}=48$, ranking candidates by $\widehat{s}_v$. The Compensation-MLP predicts 64 residual-energy bins, and the resulting compensation tail starts at $t_{\mathrm{comp}}=40$~ms. Training and evaluation used an Intel Xeon Gold 6342 CPU, an NVIDIA A100 GPU, and 512~GB of RAM.

\vspace{-0.1cm}
\subsection{Comparison methods}
\label{ssec:comparison_methods}
\vspace{-0.05cm}

We compare \textit{PathRIR} with three baselines: Pyroomacoustics (Pyroom)~\cite{pyroomacoustics}, our self-implemented full-order ISM simulator (Full-ISM), and MESH2IR~\cite{mesh2ir}. Full-ISM uses the same image-source representation and RIR accumulation backend as \textit{PathRIR}, but without pruning or compensation; it verifies that our implementation matches Pyroom under identical settings and provides a controlled full-order baseline for runtime and image-source-reduction analysis. MESH2IR is evaluated on the same test rooms.

\vspace{-0.1cm}
\subsection{Acoustic accuracy evaluation}
\label{ssec:accuracy_evaluation}
\vspace{-0.05cm}

For acoustic accuracy, we compare each simulated RIR against the Pyroomacoustics~\cite{pyroomacoustics} reference. We report cosine distance (CD) and normalized mean squared error (NMSE)~\cite{rir-former,cd}, which measure waveform shape similarity and waveform error, respectively. To evaluate decay-related properties, we also report energy-decay-curve error (EDC-Err) computed from Schroeder integration~\cite{edc}, direct-to-reverberant-ratio absolute error (DRR-Err)~\cite{drr}, and reverberation-time absolute error (RT60-Err). RT60 is estimated from the EDC using a $T_{20}$ fit~\cite{edc}. All acoustic metrics are averaged over microphones and test rooms, and lower values are better.

\begin{figure}[t]
    \centering
    \includegraphics[width=1.0\columnwidth]{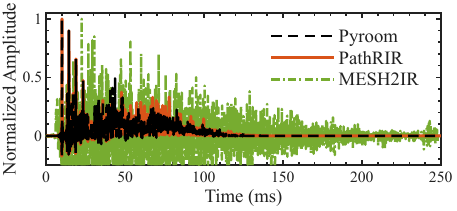}
    \vspace{-0.8cm}
    \caption{Example time-domain RIRs from Pyroom, \textit{PathRIR}, and MESH2IR for the same configuration at $O_{\max}=10$.}
    \label{fig:time_domain}
    \vspace{-0.25cm}
\end{figure}

\begin{figure}[t]
    \centering
    \includegraphics[width=1.0\columnwidth]{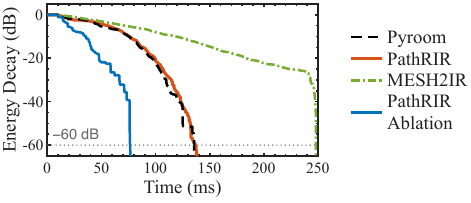}
    \vspace{-0.7cm}
    \caption{Example EDCs from Pyroom, MESH2IR, and \textit{PathRIR} with and without the Compensation-MLP at $O_{\max}=10$.}
    \label{fig:edc}
    \vspace{-0.55cm}
\end{figure}

Table~\ref{tab:results} reports acoustic accuracy against Pyroom. Full-ISM closely matches Pyroom, confirming that our Full-ISM accumulation reproduces the reference RIRs under matched image-source settings. Compared with MESH2IR~\cite{mesh2ir}, \textit{PathRIR} achieves lower errors across all five metrics, indicating better waveform and decay-related accuracy. The representative examples in Figs.~\ref{fig:time_domain} and~\ref{fig:edc} are consistent with these quantitative results: \textit{PathRIR} more closely follows the Pyroom time-domain response and EDC decay trend, whereas MESH2IR exhibits larger temporal and decay deviations. This advantage is consistent with the design of \textit{PathRIR}: neural networks are used only for path selection and late-tail compensation, while the retained reflections are accumulated using an explicit geometric simulator. In contrast, mesh-conditioned neural generation in MESH2IR does not explicitly preserve individual image-source paths or enforce the corresponding geometric constraints.

% ------------------------------------------------------------ %
\begin{table}[t]
\centering
\caption{Acoustic accuracy using Pyroom as the reference at maximum reflection order $O_{\max}=10$. Lower is better for all metrics.}
\label{tab:results}
\fontsize{9pt}{10pt}\selectfont
\setlength{\tabcolsep}{1.5pt}
\renewcommand{\arraystretch}{1.05}
\begin{tabular}{lccccc}
\toprule
\textbf{Method} &
\makecell[c]{\textbf{CD}\\[-1pt]$\downarrow$} &
\makecell[c]{\textbf{NMSE}\\[-1pt]$\downarrow$ dB} &
\makecell[c]{\textbf{EDC-Err}\\[-1pt]$\downarrow$ dB} &
\makecell[c]{\textbf{RT60-Err}\\[-1pt]$\downarrow$ ms} &
\makecell[c]{\textbf{DRR-Err}\\[-1pt]$\downarrow$ dB} \\
\midrule
MESH2IR
& 0.999 & 0.51 & 19.59 & 235.67 & 9.80 \\
Full-ISM
& $3.40{\times}10^{-7}$ & -60.64 & 0.002 & 0.05 & 0.003 \\
\textbf{\textit{PathRIR}}
& 0.141 & -5.69 & 4.69 & 36.84 & 0.54 \\
\midrule
\makecell[l]{\textit{PathRIR} w/o\\Comp. MLP}
& 0.181 & -5.09 & 18.60 & 121.12 & 2.88 \\
\bottomrule
\end{tabular}
\vspace{-0.35cm}
\end{table}
% ------------------------------------------------------------ %
% ------------------------------------------------------------ %
\begin{table}[t]
\centering
\caption{Image-source-node reduction rate $R_{\mathrm{img}}$ and
runtime speedup $S_{\mathrm{rt}}$ of \textit{PathRIR}
for $O_{\max}=1$--$10$.}
\label{tab:speed}
\fontsize{9pt}{10pt}\selectfont
\setlength{\tabcolsep}{7pt}
\renewcommand{\arraystretch}{0.95}
\begin{tabular}{@{}ccccc@{}}
\toprule
$O_{\max}$ &
$R_{\mathrm{img}}$ &
\multicolumn{3}{c}{$S_{\mathrm{rt}}\uparrow$} \\
\cmidrule(lr){3-5}
&
$\uparrow$ \% &
\makecell[c]{\textit{PathRIR}\\[-1pt]vs. Pyroom} &
\makecell[c]{\textit{PathRIR}\\[-1pt]vs. Full-ISM} &
\makecell[c]{\textit{w/o Comp. MLP}\\[-1pt]vs. Full-ISM} \\
\midrule
1  & 0.0  & 1.28   & 0.32     & 0.66     \\
2  & 0.0  & 0.60   & 0.62     & 0.83     \\
3  & 7.9  & 0.22   & 1.14     & 1.25     \\
4  & 30.8 & 0.16   & 3.75     & 3.96     \\
5  & 52.0 & 0.19   & 15.57    & 16.22    \\
6  & 67.2 & 0.51   & 73.18    & 75.62    \\
7  & 77.0 & 2.16   & 340.42   & 349.37   \\
8  & 83.4 & 10.52  & 1{,}767.16  & 1{,}807.33  \\
9  & 87.6 & 53.57  & 6{,}889.78  & 7{,}020.47  \\
10 & 90.5 & 279.39 & 42{,}133.45 & 43{,}173.95 \\
\bottomrule
\end{tabular}
\vspace{-0.35cm}
\end{table}
% ------------------------------------------------------------ %
% ------------------------------------------------------------ %
% ------------------------------------------------------------ %
% ------------------------------------------------------------ %
% ------------------------------------------------------------ %

\vspace{-0.15cm}
\subsection{Simulation efficiency evaluation}
\label{ssec:efficiency_evaluation}
\vspace{-0.1cm}

For computational efficiency, we report image-source-node reduction rate $R_{\mathrm{img}}$ and runtime speedup $S_{\mathrm{rt}}$. The reduction rate $R_{\mathrm{img}}$ measures the fraction of image-source nodes removed relative to Full-ISM. The speedup $S_{\mathrm{rt}}$ is measured separately relative to Pyroom and Full-ISM and includes Pruning-MLP inference, retained-path accumulation, and Compensation-MLP generation.

As shown in Table~\ref{tab:speed}, \textit{PathRIR} becomes faster than Full-ISM at $O_{\max}=3$ and remains faster thereafter. It is consistently faster than Pyroom from $O_{\max}=7$ onward. From $O_{\max}=4$ onward, both $R_{\mathrm{img}}$ and $S_{\mathrm{rt}}$ increase with the maximum order. At $O_{\max}=10$, \textit{PathRIR} removes $90.5\%$ of image-source nodes and achieves speedups of $279.39\times$ over Pyroom and $42{,}133.45\times$ over Full-ISM, demonstrating its increasing computational benefit for high-order simulation. At lower orders, however, the smaller number of image-source computations provides less opportunity for pruning to offset the MLP inference and compensation overhead.

\vspace{-0.2cm}
\subsection{Ablation study}
\label{ssec:ablation}
\vspace{-0.1cm}

We evaluate \textit{PathRIR} without the Compensation-MLP, as reported in the last row of Table~\ref{tab:results} and the last column of Table~\ref{tab:speed}. The pruning-only variant is faster at every tested order, but its advantage generally narrows as $O_{\max}$ increases. At $O_{\max}=10$, it is approximately $2.5\%$ faster than the full \textit{PathRIR} ($43{,}173.95\times$ versus $42{,}133.45\times$ over Full-ISM). Removing the Compensation-MLP, however, degrades all five acoustic metrics. CD increases from 0.141 to 0.181, while NMSE worsens from -5.69~dB to -5.09~dB. The degradation is greater for decay-related metrics: EDC-Err increases from 4.69~dB to 18.60~dB, RT60-Err from 36.84~ms to 121.12~ms, and DRR-Err from 0.54~dB to 2.88~dB. Figure~\ref{fig:edc} shows that the pruning-only EDC decays much faster than Pyroom because pruning removes many late-tail image sources. The Compensation-MLP mitigates this by adding an envelope-matched compensation tail to the pruned RIR. Although this tail does not recover the exact residual waveform, it helps \textit{PathRIR} follow the Pyroom decay trend and reduces waveform- and decay-related errors at modest runtime cost.

% ============================================================= %
% ============================================================= %
% ============================================================= %

\vspace{-0.35cm}
\section{Conclusion}
\label{sec:conclusion}
\vspace{-0.15cm}

We presented \textit{PathRIR}, a physics-guided framework for accelerating ISM-based RIR simulation while preserving the image-source structure of geometric acoustics. Instead of replacing the RIR simulator with a neural generator, \textit{PathRIR} uses a Pruning-MLP to retain acoustically important paths and a lightweight Compensation-MLP to restore the missing late-tail energy envelope. Experiments on irregular 3D rooms show that its computational gains grow rapidly with $O_{\max}$. At $O_{\max}=10$, \textit{PathRIR} removes $90.5\%$ of image-source nodes and achieves speedups of $279.39\times$ over Pyroom and $42{,}133.45\times$ over Full-ISM, while yielding lower errors than MESH2IR across all five acoustic metrics. The ablation shows that the Compensation-MLP improves all five metrics, with particularly large gains in decay-related accuracy, at modest runtime cost. Future work will explore ray-tracing or hybrid ISM/ray-tracing backends to better capture mid-to-late reverberation.

% ============================================================= %
% ============================================================= %
% ============================================================= %

\bibliographystyle{IEEEbib}
\bibliography{refs}

\end{document}